\documentclass[a4paper,11pt]{article}
\pdfoutput=1 

\usepackage{jcappub} 
\usepackage[T1]{fontenc} 
\usepackage{orcidlink}
\usepackage{subcaption}

\def\b{\begin{equation}}
 \def\e{\end{equation}}

\newcommand{\dif}{\mathrm{d}}
\newcommand{\bq}{\begin{eqnarray*}}
\newcommand{\eq}{\end{eqnarray*}}
\newcommand{\beq}{\begin{eqnarray}}
\newcommand{\enq}{\end{eqnarray}}

\title{\boldmath Large primordial non-Gaussianity from transient turns in  Higgs-$R^2$ inflation}

\author[a]{Flavio Pineda\orcidlink{0000-0002-1265-5229}}
\author[a]{Luis O. Pimentel\orcidlink{0000-0002-3614-7237}}

\affiliation[a]{Departamento de Física, Universidad Autónoma Metropolitana Iztapalapa,\\
Av. San Rafael Atlixco No.~186, Col. Vicentina, 09340 CDMX, Mexico}

\emailAdd{fpineda@xanum.uam.mx}
\emailAdd{lopr@xanum.uam.mx}

\abstract{
We investigate the generation of primordial non-Gaussianities in multifield Higgs--$R^2$ inflation, focusing on the effects of transient turning trajectories in the hyperbolic field space manifold. We compute the full bispectrum without relying on slow-roll or local approximations and follow the complete superhorizon evolution of curvature and isocurvature perturbations. We show that transient turns efficiently transfer isocurvature fluctuations into the adiabatic sector, generating sizeable local non-Gaussianities. For a benchmark Higgs nonminimal coupling $\xi_h = 0.1$ and quartic coupling $\lambda = 10^{-10}$, we obtain $f_{\rm NL}^{\rm loc}\simeq -17.7$. As the Higgs nonminimal coupling increases, the turning rate is progressively suppressed and the model approaches the effective single-field attractor, recovering the Maldacena consistency relation $f_{\rm NL}\rightarrow 0.0159$. Comparing our predictions with current CMB constraints, we find that primordial non-Gaussianity provides a sensitive probe of the Higgs nonminimal coupling and can significantly restrict the viable parameter space of the model. 
}

\begin{document}
\maketitle
\flushbottom

\section{Introduction}

Cosmic inflation remains the leading paradigm of the early universe, successfully  explaining and predicting the origin of the primordial perturbations that seed both 
large-scale structure formation and the anisotropies of the cosmic microwave  background (CMB)~\cite{Guth1981, Linde1982}. Standard single-field inflation  predicts a nearly scale-invariant primordial power spectrum of Gaussian and  adiabatic perturbations~\cite{Stewart1993, weinberg2003}, which is highly  compatible with the latest high-precision CMB observations~\cite{Planck2018_inflation, AtacamaDR62025}. However, establishing a well-motivated inflationary scenario  within high-energy physics frameworks remains a major challenge for modern cosmology. Realistic inflationary models inspired by high-energy physics often 
require more than one scalar field to describe the dynamics of the early 
universe~\cite{Lyth1999, Kaiser2010a, Gorbunov2019}. The key distinction from  single-field models is that perturbations are not necessarily purely adiabatic; multifield inflationary models generally produce isocurvature modes that can survive past the end of inflation~\cite{starobinsky1985, Bartolo2001, Wands2002, Huston2012, Kaiser2014, Schutz2014}, or generate features in the primordial curvature spectrum through the coupling of adiabatic and isocurvature modes on super-horizon scales~\cite{Braglia2020}. Understanding this coupling and the overall perturbation dynamics has a direct impact on the primordial power spectrum, 
on cosmological observables such as $n_s$ and $r$~\cite{Bartolo2001, Wands2002, Achucarro2011}, and on the production of primordial non-Gaussianities~\cite{Seery2005, Langlois2008, Elliston2012, Gao2008, Gong2011}.

Primordial non-Gaussianities generated during the inflationary epoch have 
become a standard and powerful tool for probing the physics of the early universe.  They offer unique observational signatures that discriminate among inflationary  models by revealing the field content and interactions present in the early universe. On the one hand, single-field inflation features a distinctive property regarding non-Gaussianity: the bispectrum is highly suppressed in the squeezed limit, where a long-wavelength mode couples with two short-wavelength modes~\cite{PaoloCreminelli2004, 
Ganc2010}. For any single-field inflationary model with a canonical kinetic 
term, the non-linearity parameter $f_\text{NL}$ in this limit is given 
by~\cite{Maldacena2003}

\begin{equation}
    f_\text{NL} = \dfrac{5}{12}(1-n_s)\,,
    \label{maldacena result}
\end{equation}
where $n_s = 0.968 \pm 0.003$ ($68\%$ C.L.) is the scalar spectral index 
measured by recent CMB observations, including Planck 2018 and ACT-DR6~\cite{Planck2018_inflation,AtacamaDR62025}. The result \eqref{maldacena result} is known as the Maldacena consistency relation; the non-linear parameter is highly suppressed in the squeezed limit ($k_3\ll k_1\simeq k_2$) $f_\text{NL} \ll 1$. On the other hand, in the context of multifield inflation, primordial non-Gaussianities — and in particular the amplitude of the local bispectrum, $f_\text{NL}^\text{loc}$ — represent a primary target for next-generation CMB observations. Notably, a detection of local non-Gaussianity of order $\mathcal{O}(1-10)$  would rule out all single-field inflation models with canonical kinetic term. The complex dynamics inherent to multifield scenarios can generate significantly larger non-Gaussianities across both large and small scales (for current observational bounds, see~\cite{Planck2018_non_gaussianity}). A systematic characterization of when this occurs was provided by Ref.~\cite{Byrnes2008}, which identifies four conditions for generating large primordial non-Gaussianity:

\begin{enumerate}
    \item Multifield inflation.
    \item Noncanonical kinetic terms.
    \item Slow-roll violation.
    \item Initial state different from the Bunch-Davies vacuum
\end{enumerate}

In multifield inflation models, the first three of these conditions are generally satisfied. In particular, a rapid transient turn $\eta_\perp$ efficiently transfers isocurvature fluctuations into the adiabatic curvature sector on super-horizon scales. This mechanism generates distinct primordial non-Gaussianities, producing large local or non-local shapes because fluctuations continue to evolve outside the horizon. A multifield inflationary model with a hyperbolic field space, whose dynamics are characterized by strongly non-geodesic motion, can produce hyper non-Gaussianities \cite{Fumagalli_2019}. A strong non-geodesic motion can drive rapid and transient turns in field space \cite{RenauxPetel2022, Wolters2024, Anguelova2024, Basiouris2025, Aragam2022, Achcarro2020, Bjorkmo2019, Chakraborty2020}, generating non-Gaussian signals \cite{Fumagalli_2019, Iarygina2024, Aoki2026qea, Iacconi2023} and imprinting the background dynamics onto the production of PBHs \cite{Fumagalli2023}.

Among the many multifield inflationary models in the literature \cite{Martin2024}, the Higgs-$R^2$ model stands out as a theoretically well-motivated and predictive framework~\cite{Ema2017,He2018,Wang2017, Gundhi2020}. This model is a combination of the Higgs inflation~\cite{Bezrukov2008} with the Starobinsky model~\cite{starobinsky1980} as a result of attempts to address the strong coupling problem of Higgs inflation~\cite{Giudice_2011, Salvio2015,Bezrukov2008}. The presence of the $R^2$ term in the Higgs inflation action allows to push the strong-coupling scale in all sectors of the model up to the Planck scale $M_p$~\cite{Ema2017,Gorbunov2019}. The result is a two-scalar field model of inflation with noncanonical kinetic term. This structure makes the model an ideal well-motivated high-energy physics model of inflation, with a rich phenomenology related to primordial perturbations. Previous studies of the Higgs-$R^2$ model have focused primarily on the background dynamics~\cite{Ema2017,He2018,Gundhi2020}, primordial perturbations and primordial features~\cite{Wang2017,Pineda2025}, gravitational waves and primordial black hole production~\cite{Cheong2021,Cheong_2022, Kim2025}, preheating after inflation~\cite{BEZRUKOV2019,He2019,He2021,He2021tachionic} and gravitational particle production~\cite{Pineda2025a}. However, a systematic analysis of primordial non-Gaussianities and their dependence on the nonminimal Higgs coupling is still lacking.

In this work, we compute the primordial non-Gaussianity generated in the Higgs-$R^2$ model of inflation, taking fully into account its multifield dynamics. We focus on the region of parameter space characterized by a nonminimal Higgs coupling of order $\xi_h \sim \mathcal{O}(0.1)$ and a quartic self-coupling $\lambda \sim \mathcal{O}(10^{-10})$, which is compatible with the observed amplitude of primordial scalar perturbations. Using the $\delta N$ formalism, we derive analytical expressions for the non-linearity parameter $f_\text{NL}$ and investigate the resulting bispectrum, with particular emphasis on the local shape. We find that the model can generate large primordial non-Gaussianities within the parameter region under consideration, provided that suitable initial conditions are chosen. In particular, sizeable values of $f_\text{NL}^{\rm loc}$ arise when the inflationary trajectory begins near the top of the natural ridge structure of the potential, requiring a degree of fine-tuning in the initial field configuration~\cite{Yokoyama2007,Peterson2011,Peterson2011a}. However, as the nonminimal coupling increases, the non-Gaussian signal is rapidly suppressed. For $\xi_h \gtrsim 0.12$, the non-linearity parameter approaches the single-field consistency relation prediction, indicating a transition from a genuinely multifield regime with efficient isocurvature-to-curvature transfer to an effectively single-field attractor.

The remainder of this paper is organized as follows. In Sec.~\ref{sec2} we present a brief review of the Higgs-$R^2$ model focused on the covariant description. In Sec.~\ref{sec3}, we present the covariant approach for perturbations and the $\delta N$ formalism to evaluate the 3-point correlation function $\langle \zeta \zeta \zeta \rangle$. In Sec.~\ref{sec4}, we present the numerical results for the non-linear parameter and the shape of the bispectrum. Lastly, in Sec.~\ref{sec5} we present the conclusions of our results. In the following, we work with a spatially flat FLRW background,

\begin{equation}
\dif s^2 = -\dif t^2 + a^2(t)\,(\dif x^2 + \dif y^2 + \dif z^2),
\label{FLRW metric}
\end{equation}
and use natural units with $c = \hbar = G = M_p = 1$.


\section{The Higgs--$R^2$ model: a brief review}    
\label{sec2}

We consider the model of inflation which consist in the SM Higgs boson with nonminimal coupling $\xi_h$ and a quadratic term in the Ricci curvature $R$. This combination naturally arises when one seeks a UV-complete inflationary model consistent with the Standard Model of particle physics.
There are several studies of this model in the literature, we refer the reader to the pioneering works \cite{Salvio2015, Ema2017, Gorbunov2019, Gundhi2020} and the references therein. We work in unitary gauge and neglect gauge interactions during inflation. In the Jordan frame, the action is given by 

\begin{equation}
     S_J[g_{\mu\nu}\,,h] = \int d^4x \sqrt{-g_J} \left[ \frac{1}{2}(1 + \xi_h h^2)R_J + \xi_s R_J^2  - \frac{1}{2}g_J^{\mu\nu}(\partial_\mu h)(\partial_\nu h) - \frac{\lambda}{4}h^4 \right] \,,
    \label{Higgs-R2 action}
\end{equation}
 where the variables defined in the Jordan frame are denoted by the subscript $J$. The field $h$ represents the Higgs field in the unitary gauge with a nonminimal coupling $\xi_h$, which is necessary for $h$ to drive inflation, while $\lambda$ is the quartic coupling constant.
 This model represents a natural UV extension of Higgs inflation in order to solve the strong-coupling problem in the Higgs-like inflation \cite{Gorbunov2019, Bezrukov2008}. The quadratic term $R^2_J$ represents a new scalar degree of freedom present during the inflationary epoch. Thus, this model describes two scalar fields that contribute to the inflationary dynamics. We can perform a conformal transformation $g_{\mu\nu}\to \Omega^2(x)g_{\mu\nu}$ to remove the quadratic term in $R_J$ and the nonminimal coupling constant $\xi_h$. The conformal transformation is equivalent to introducing a new scalar field called scalaron $\phi$, i.e.,
$\Omega^2(x) = e^{-\alpha\phi}$, where $\alpha = \sqrt{2/3}$ in Planck units. This allows us to rewrite the action \eqref{Higgs-R2 action} in terms of two scalar fields $(\phi\,,h)$, with a scalar potential $V(\phi\,,h)$ in the Einstein frame

\begin{equation}
       S_E = \int \dif^4 x\sqrt{-g}\left[\dfrac{1}{2}R - \dfrac{1}{2}g^{\mu\nu}(\partial_\mu \phi)(\partial_\nu \phi) - \dfrac{1}{2}e^{-\alpha\phi}g^{\mu\nu}(\partial_\mu h)(\partial_\nu h) - V(\phi\,,h)\right]\,,
    \label{Einstein frame action}
\end{equation}
where the scalar potential $V(\phi\,,h)$ depends on both fields in the Einstein frame and is defined as

\beq
V(\phi\,,h) = e^{-2\alpha\phi}\left[\dfrac{1}{16\xi_s}\left(e^{\alpha\phi} - 1 - \xi_h\,h^2 \right)^2 + \dfrac{\lambda}{4}\,h^4\right]\,.
    \label{potential Einstein}
\enq

In the Einstein frame, both fields are coupled via the nonminimal kinetic term $e^{-\alpha\phi}$. This introduces a non-flat two-dimensional field space metric $G_{IJ}=\text{diag}(1\,, e^{-\alpha\phi})$ allowing us to employ the covariant formalism of multifield inflation \cite{Gong2011,Langlois2008, Achucarro2011}. The associated Christoffel symbols and field space curvature are non-vanishing, leading to geometrical contributions in the perturbation dynamics. In fact, the non-zero Christoffel symbols and the field space scalar curvature are given by 

\beq
\Gamma^\phi{}_{h h} = \dfrac{\alpha}{2}e^{-\alpha\phi}\,,\quad \Gamma^h{}_{\phi h} = \Gamma^h{}_{h \phi} = -\dfrac{\alpha}{2}\,,\quad R_\text{fs} = -\dfrac{1}{3}\,.
\label{Christoffel}
\enq

The negative value of $R_\text{fs}$ ensure that the field space geometry is hyperbolic. Moreover, the negative curvature of field space enhances geometrical forces orthogonal to the background trajectory, leading to transient turning episodes whose duration and amplitude depend on the initial displacement from the ridge, within the parameter region considered. In this context, the noncanonical kinetic term on the Higgs field $h$ can be understood as a geometric interaction between fields, having an impact on the isocurvature mass and on the shape of the bispectrum \cite{Kaiser2013, Elliston2012}.

The parameter $\alpha$ is fixed by the conformal transformation from the Jordan to the Einstein frame, yielding $\alpha = \sqrt{2/3}$. This parameter is responsible for the kinetic coupling between the Higgs field $h$ and the scalaron $\phi$, meaning it cannot be varied without fundamentally altering the gravity sector of the theory. In addition, the model has three parameters: $\lambda$, $\xi_h$, and $\xi_s$. The constant $\xi_s$ is related to the scalaron mass $M$ as $\xi_s = 1/3M^2$, where $M = 1.3\times 10^{-5}$ in Planck units. The quartic coupling constant $\lambda$ takes a standard low-energy value of $\lambda = 0.13$, which is dictated by the mass of the Higgs boson via $\lambda = 3m_h^2/v^2$, where $v = 246\,\mathrm{GeV}$ is the vacuum expectation value. However, since the Higgs boson mass depends on the energy scale and the top quark mass, the renormalization group equations allow for a quartic coupling constant of order $\mathcal{O}(10^{-10})$ at inflationary scales \cite{Bezrukov2011, Lebedev2011, Burgess2010, Salvio2015, Gundhi2020, Cheong_2022}. 


\subsection{Background dynamics}

The multifield dynamics of the model have a direct impact on the perturbation analysis and the resulting phenomenology. As established in  \cite{Pineda2025}, for $\xi_h \sim \mathcal{O}(10^{-1})$, $\lambda = 10^{-10}$ and $\xi_s = 4\times 10^8$, multifield effects become relevant, there are isocurvature modes that leave their mark on the primordial curvature spectrum $\mathcal{P}_\zeta(k)$, which has a direct impact on the CMB observables. Thus, it is natural to expect that in this parameter space, the isocurvature modes can generate a non-Gaussian signal sourced by primordial fluctuations. To investigate this, we track the background evolution by solving the covariant equations of motion

\begin{equation}
    D_t \dot{\varphi}^I + 3H \dot{\varphi}^I + G^{IJ} V_J = 0\,,
    \label{background equation}
\end{equation}
where $D_t \dot{\varphi}^I = \ddot{\varphi}^I + \Gamma^I_{JK}\dot{\varphi}^J\dot{\varphi}^K$, dots specify cosmic time derivatives and $I=\{\phi,h\}$. The Friedmann equations read

\begin{equation}
    3H^2 = \frac{1}{2}\dot{\sigma}^2 + V(\phi,h)\,,\quad \dot{H} = -\dfrac{1}{2}\,\dot{\sigma}^2\,,
    \label{Friedmann equation}
\end{equation}
where $\dot{\sigma}^2 = G_{IJ}\dot{\varphi}^I\dot{\varphi}^J = \dot{\phi}^2 + e^{-\alpha\phi}\,\dot{h}^2$. As in single-field inflation, we can define the slow-roll parameter $\epsilon_H$ as

\beq 
    \epsilon_H = -\dfrac{\dot{H}}{H^2} = \dfrac{\dot{\sigma}^2}{2H^2}\,,
    \label{slow-roll parameter}
\enq
which can be expressed in terms of the matrix $\epsilon^{IJ}$ 

\beq
\epsilon^{IJ} = \dfrac{\dot{\phi}^I\,\dot{\phi}^J}{2 H^2}\,,\quad \epsilon_H = \text{Tr}\,\epsilon^{IJ}
\label{slow-roll matrix}
\enq

To characterize the background trajectory, it is convenient to introduce the adiabatic and entropic unit vectors $e_\sigma^I$ and $e_s^I$ defined along and orthogonal to the background trajectory, respectively

\beq
    e_\sigma^I = \dfrac{\dot{\phi}^I}{\dot{\sigma}} \,,\quad e_s^I = \sqrt{\det G}\, \varepsilon^{IJ}\,e_{J \sigma}\,,  
    \label{unit vectors}
\enq
where $\varepsilon^{IJ}$ is the totally antisymmetric 2-dimensional Levi-Civita tensor and $\det G$ is the determinant of the field space metric $G_{IJ}$. The basis $\mathcal{B} = \{e^I_\sigma\,,e^I_s\}$ is known as kinematic basis, and fulfills the conditions

\beq
    G_{IJ} = e_I{}^\alpha e_J{}^\beta \delta_{\alpha\beta}\,,\quad  G_{IJ}\,e^I{}_\alpha e^J{}_\beta = \delta_{\alpha\beta}
    \label{kinematic basis}
\enq

The covariant transport of these vectors is represented by the directional derivative $D_t$ such that 

\beq 
    D_t e_\sigma^I= -\eta_\perp\, e_s^I\,,\quad D_t e_s^I = \eta_\perp\,e_\sigma^I\,,
\enq 
where $\eta_\perp$ is the dimensionless turning rate, defined as

\begin{equation}
    \eta_\perp \equiv  \frac{V_{N}}{H \dot{\sigma}} = \dfrac{\dot{\theta}}{H}\,,
    \label{turnig rate}
\end{equation}
where $V_{N} = e_s^I V_I = e_s^I\partial_I V$ is the projection of the gradient orthogonal to the trajectory and $\dot{\theta} = V_N/\dot{\sigma}$. For the parameter region $\xi_h \sim 0.1$ and $\lambda = 10^{-10}$, trajectories starting near the ridge experience a transient enhancement of the turning rate $\eta_\perp$, while trajectories closer to the valley rapidly align with the adiabatic direction. As we will show, this transient 
turn plays a crucial role in the generation of non-Gaussianity.

\begin{figure}[t!]
    \centering
    \includegraphics[width = 0.48\textwidth]{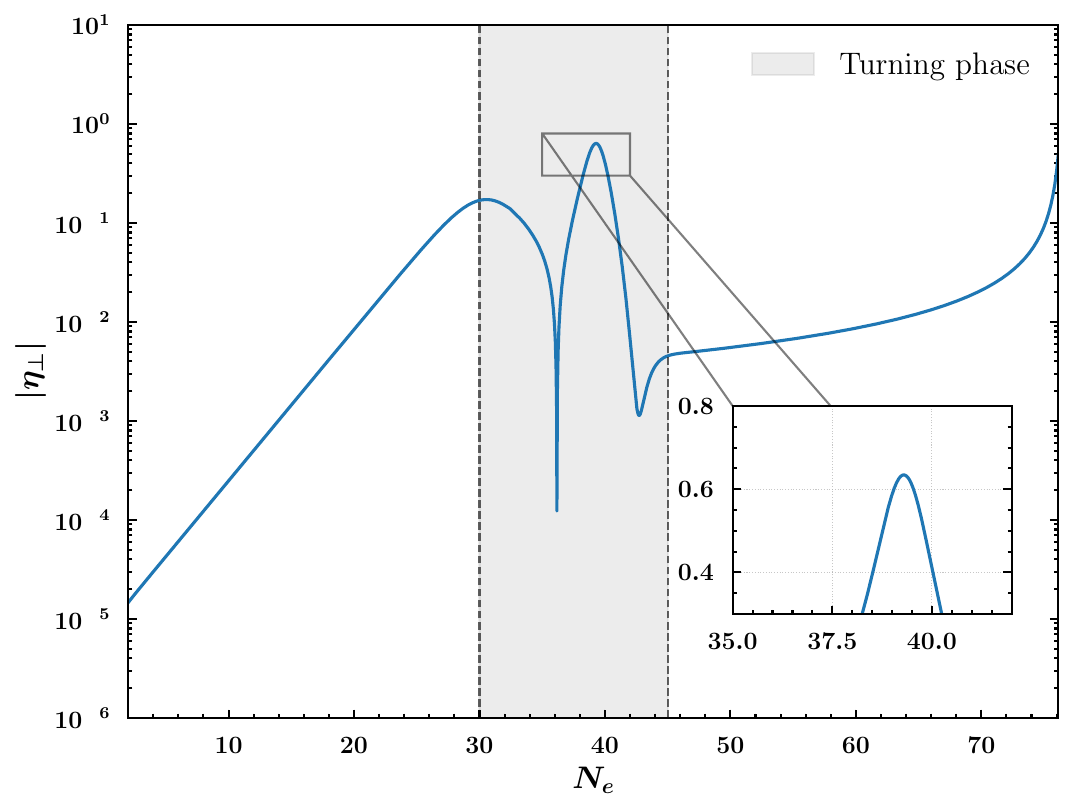}\quad
    \includegraphics[width = 0.48\textwidth]{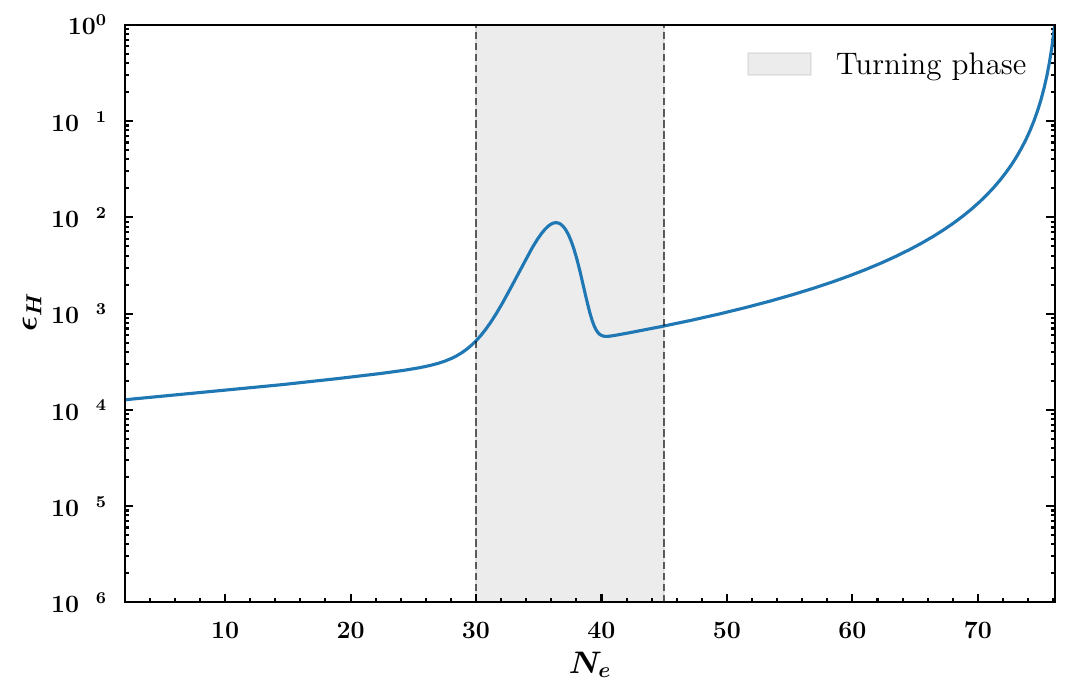}
    \caption{The turning rate parameter (left panel), and slow-roll parameter $\epsilon_H$ (right panel) as a function of the number of e-folds $N_e$ for the initial conditions $h_0 = 10^{-5}$ and $\phi_0 = 5.7$ in Planck units, and the parameter values $\xi_h = 0.1\,,\lambda = 10^{-10}\,,\xi_s = 4\times 10^8$. The maximum value of $\eta_\perp$ is $\max(\eta_\perp) = 0.634327 $, while the gray region is the e-fold interval over which the transient turn occurs. The end of inflation takes place at $N_\text{end} \approx 76.12$}
    \label{fig1}
\end{figure}

In Fig.~\ref{fig1}, we show the behavior of the turning rate \eqref{turnig rate} and the slow-roll parameter $\epsilon_H$ as a function of the e-folds $N_e$ for suitable initial conditions. In this model, the associated multifield effects are related to the initial position of the Higgs field; it is necessary for the background trajectory to remain at the ridge long enough for the fall into one of the valleys to occur in the time interval in which the relevant modes cross the horizon \cite{Gundhi2020}. For trajectories starting near the ridge, the combined effect of the potential gradient and the hyperbolic field space geometry induces a transient departure from geodesic motion. This manifests as a sharp peak in the turning rate $\eta_\perp$, signaling a rapid rotation of the adiabatic direction. After the turn, the trajectory aligns with the valley and the bending rapidly decays. The mild feature in $\epsilon_H$ reflects a temporary exchange between potential and kinetic energy during the turn.

\begin{figure}[t!]
    \centering
   \includegraphics[width = 1\textwidth]{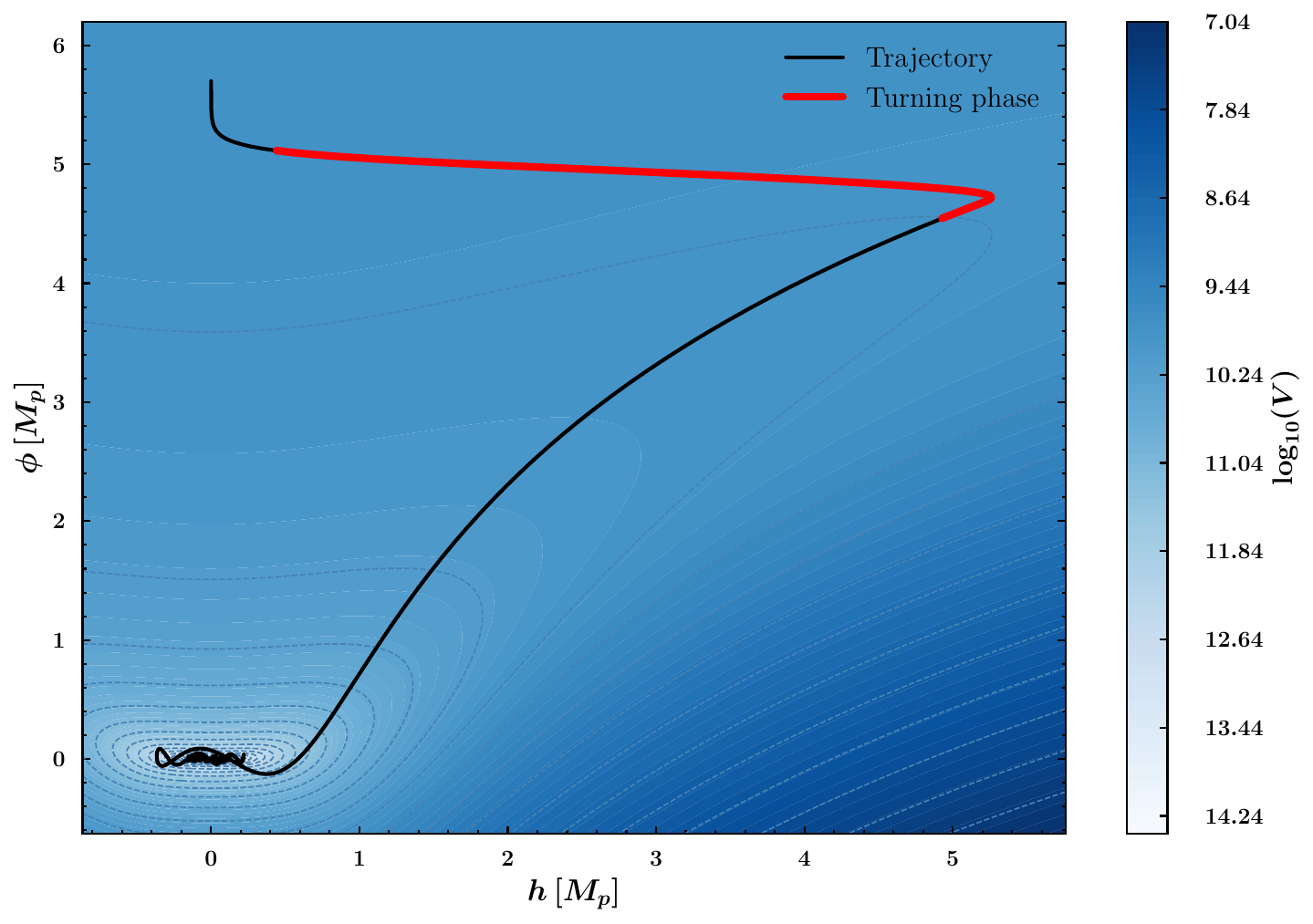}
    \caption{Evolution of the fields $\phi$, $h$ over the scalar potential $V(\phi\,,h)$, Eq. \eqref{potential Einstein}. The background color map represents $\log_{10}(V)$, with lighter colors denoting the minimum of the potential valley. The trajectory (black) is strongly deflected by the steep walls of the potential, resulting in a pronounced turning phase (red) where the turning rate $\eta_\perp$ peaks. This kinematic deviation from a straight path is responsible for the coupling between adiabatic and isocurvature modes.}
    \label{fig1.1}
\end{figure}

\subsection{Effective single-field reduction}

As mentioned previously, the model features three primary parameters: $\lambda$, $\xi_h$, and $\xi_s$. According to \cite{Ema2017}, there are three distinct regions in parameter space that yield successful inflation in the large-field regime ($\phi \gg M_p$):

\begin{enumerate}
    \item $\xi_h > 0$, $\xi_s > 0$, $\lambda > 0$
    \item $\xi_h < 0$, $\xi_s > 0$, $\lambda > 0$
    \item $\xi_h < 0$, $\xi_s > 0$, $\lambda < 0$
\end{enumerate}

In all three cases, the inflationary dynamics effectively reduces to a single-field scenario for $\xi_h \gtrsim 1$, as the isocurvature mode becomes sufficiently heavy to be integrated out. This dynamical reduction can be well understood via the valley approximation, as extensively discussed in the foundational literature of the model \cite{Ema2017, He2018, Gundhi2020}. 

The first case corresponds to a trajectory driven by a mixture of the Higgs field $h$ and the scalaron $\phi$. The parameter hierarchy plays a crucial role in this case: if $\xi_s \gg \xi_h^2/\lambda$, inflation is strictly of the $R^2$-type, whereas in the opposite limit, it is predominantly driven by the Higgs field $h$. In these regimes, the effective inflaton potential along the valley is given by
\begin{equation}
    V_\text{eff}(\phi, h_0(\phi)) = \frac{\lambda}{4(\xi_h^2 + 4\lambda\xi_s)}(1 - e^{-\alpha\phi})^2 \, .
    \label{eq:effective_potential}
\end{equation}

However, if no strict hierarchy exists among the parameters (i.e $\xi_h \sim \lambda \xi_s$), both $\phi$ and $h$ actively contribute to the background inflationary trajectory. In the second case, inflation is driven solely by the scalaron $\phi$, while the third case corresponds to inflation within a metastable electroweak vacuum. 
Despite this mixed background evolution, for large non-minimal couplings ($\xi_h \gtrsim 1$), the generation of primordial perturbations is entirely dominated by the adiabatic mode. This specific regime is known as the standard parameter space of the Higgs-$R^2$ model, where observable multifield effects (such as isocurvature sourcing and large non-Gaussianities) are known to be fundamentally absent.


\section{Analysis of perturbations}
\label{sec3}

The evolution of the background plays an important role for the primordial perturbations and for the generation of non-Gaussianities. In this section, we introduce the adiabatic and isocurvature perturbations in a covariant form, and we briefly describe the $\delta N$ formalism to multifield inflationary models for calculating the non-Gaussianity signal of the Higgs-$R^2$ model.

\subsection{Covariant perturbations}

In order to study the non-linear primordial perturbations, we adopt the covariant formalism developed by Gong and Tanaka \cite{Gong2011}. The main idea is use the covariant description of the field space to perturbatively expand the field perturbations $\delta \phi^I$ in terms of the geometric field space quantitites, particularly in terms of the tangent-space vector $Q^I$. 
We set the background trajectory as $\phi_0^I(t)$ and the perturbated field $\phi^I(x)$, where $x$ is a space-time point. We also define the field perturbation $\delta\phi^I$ as

 \beq 
    \delta \phi^I(t\,,\mathbf{r}) = \phi^I(t\,,\mathbf{r}) - \phi_0^I(t)\,,
\enq 

These fields are linked by a unique geodesic parametrized by a parameter $\lambda$, such that at $\lambda =0$ corresponds to the unperturbated background variable $\phi_0^I$ and $\lambda = 1$ corresponds to the perturbed variable $\delta\phi^I + \phi_0^I$.  The covariant perturbation $Q^I$ is defined as the tangent vector $Q^I = \left.\dif \phi^I /\dif \lambda\right|_{\lambda =0}$. Then we can perform a Taylor series of the vector field $\phi^I$ around $\lambda = 0$

\beq
    \phi^I = \bar{\phi}^I + \left.\dfrac{\dif \phi^I}{\dif \lambda}\right|_{\lambda = 0}  + \dfrac{1}{2!}\left.\dfrac{\dif^2 \phi^I}{\dif \lambda^2}\right|_{\lambda = 0}  + \dfrac{1}{3!}\left.\dfrac{\dif^3 \phi^I}{\dif \lambda^3}\right|_{\lambda = 0}  + \cdots\,.
    \label{taylor series}
\enq
which allows us to rewrite the field perturbation $\delta\phi^I$ in term of $Q^I$ and the Christoffel symbols $\Gamma^I{}_{JK}$ of the field space

\begin{equation}
       \delta\phi^I = \phi^I-\bar{\phi}^I = Q^I - \dfrac{1}{2}\Gamma^I{}_{JK}\,Q^
    JQ^K 
     + \dfrac{1}{6}(\Gamma^I{}_{LM}\Gamma^M{}_{JK}-\Gamma^I{}_{JK,L})\, Q^J Q^K Q^L + \cdots\,,
    \label{inflationary perturbation}
\end{equation}
where we have used the geodesic equation 

\beq 
    D_t \dot{\phi}^I= \dfrac{\dif^2 \phi^I}{\dif t^2} + \Gamma^I{}_{JK}\dfrac{\dif \dot{\phi}^J}{\dif t}\dfrac{\dif \dot{\phi}^K}{\dif t}=0\,\,.
    \label{geodesic equation}
\enq 

Since $Q^I$ lies in the tangent direction $e_\sigma^I$ of the field space, any tensorial quantity can be perturbatively expanded in terms of $Q^I$, and it will be manifestly covariant. For example, the expansion of the scalar potential $V(\phi^I)$ in terms of $Q^I$ reads as

\begin{equation}
      V(\phi^I) = V_0 + Q^I V_I + \dfrac{1}{2!}\,Q^I Q^J \nabla_I V_J 
    + \dfrac{1}{3!}Q^IQ^JQ^K \nabla_I \nabla_J V_K + \cdots\,,
    \label{potential perturbation}
\end{equation}
where $\nabla_I$ is the covariant derivative of the field space given by $\nabla_I A^J = \partial _I A^J + \Gamma^J_{IK} A^K $ for any vector field $A^I$ of the field space. 

\subsection{Curvature and isocurvature perturbations}

In our previous paper \cite{Pineda2025}, we carried out an exhaustive analysis of the primordial perturbations in this model. In this section, we briefly summarize the most relevant aspects of the perturbation dynamics at linear order. 
The comoving gauge is defined by $\gamma_{ij}^\mathrm{com} = a^2(t)e^{2\zeta}\delta_{ij}$ and $e_{\sigma\,I} Q^I = 0$, where $\zeta$ is the gauge-invariant comoving curvature perturbation \cite{Mukhanov1992}. The main feature of this gauge is that the adiabatic component of $Q^I$ vanishes, leaving only the isocurvature part $Q_s = e_{s\,I} Q^I$. Consequently, the dynamical variables are $\zeta$ and $Q_s$. The action at second order in these variables reads

\begin{multline}
     S^{(2)} = \int \dif^4 x\, a^3\left[ \epsilon_H\left(\dot{\zeta}^2-a^{-2}\delta^{ij}\partial_i\zeta\partial_j\zeta\right) 
     + \dfrac{1}{2}\left(\dot{Q}_s^2-a^{-2}\delta^{ij}\partial_iQ_s \partial_j Q_s- m_\mathrm{iso}^2Q_s^2\right)\right.\\
     \left.+ 2H \eta_\perp \sqrt{2\epsilon_H}\, Q_s \dot{\zeta} \right]\,,
        \label{second order action}
\end{multline}
where $\epsilon_H$ is the slow-roll parameter given by \eqref{slow-roll parameter}, and $m^2_\text{iso}$ is the effective mass of the isocurvature mode $Q_s$, given by

\beq
    m_\mathrm{iso}^2 = e_s^I e_s^J \nabla_I V_J + H^2\epsilon_H\,R_\text{fs}-(H\eta_\perp)^2\,,
    \label{iso mass}
\enq
where $R_\text{fs} $ is the field space Ricci scalar in units of $M_p$, defined in \eqref{Christoffel}. In \cite{Pineda2025}, we showed that the nonminimal coupling $\xi_h$ controls the multifield dynamics, with a direct impact on inflationary observables such as the isocurvature fraction $\beta_\text{iso}$, the tensor-to-scalar ratio $r$, and the CMB angular power spectra. In the weak-coupling regime ($\xi_h \ll 1$), isocurvature modes survive until the end of inflation, leaving a remnant of about $1\%$. However, the turning rate parameter remains of order $10^{-15}$ throughout inflation, implying that $\zeta$ and $Q_s$ evolve independently. Conversely, the intermediate-coupling regime ($\xi_h \sim \mathcal{O}(0.1)$) induces an energy transfer from the isocurvature perturbation $Q_s$ to the adiabatic perturbation $\zeta_k$ on super-horizon scales via the source term

\beq 
    \dot{\zeta}_k \simeq -\dfrac{\sqrt{2}H\eta_\perp}{\sqrt{\epsilon_H}}Q_s\,.
\enq

In this regime, the turning rate \eqref{turnig rate} exhibits a transient peak (see Fig.~\ref{fig1}) that sources the long-wavelength evolution, modifying the primordial curvature power spectrum. This super-horizon dynamics is primarily responsible for the generation of a non-Gaussian signal, making it the ideal scenario to apply the $\delta N$ formalism. Regarding the stability of the multifield dynamics, it is instructive to briefly comment on the mass scales of the perturbations. The effective mass of the isocurvature perturbations, $m_\text{iso}^2$, was analytically derived in the valley approximation in our previous work \cite{Pineda2025} (see Sec. III). While the negatively curved field space ($R_\text{fs} = -1/3$) and the transient turning rate ($\eta_\perp$) introduce negative contributions to \eqref{iso mass}, the dominant term arises from the potential's curvature in the transverse direction ($V_{ss} = e_s^I e_s^J \nabla_I V_J $). Because the trajectory is securely confined within a stable valley of the potential (defined along the adiabatic direction $e_{\sigma}^I$), where the transverse gradient vanishes $V_s \to 0$, which implies $\eta_\perp \to 0$. Even though both fields $\phi$ and $h$ remain dynamically active during the initial phase, this positive transverse curvature ensures that $m_\text{iso}^2 > 0$ throughout the entire evolution. This positive effective mass guarantees that the multifield dynamics remain stable against tachyonic instabilities during the inflationary epoch (for general discussions on geometrical destabilization and tachyonic instabilities during multifield inflation, see e.g., \cite{Renaux_Petel_2016}).

We emphasize that this stability strictly applies to the generation of the primordial signal with constant effective couplings at the inflationary scale. However, if the full renormalization group running of the parameters is explicitly taken into account, tachyonic instabilities can also be induced during the inflationary phase, leading to phenomena such as primordial black hole formation \cite{Cheong_2022}. Furthermore, even without RG running, after inflation ends, as the fields oscillate around the global minimum and climb up the $\phi > 0$ region with $h \simeq 0$, the Higgs field can indeed experience a tachyonic instability. Detailed analyses of this post-inflationary tachyonic preheating in the Higgs-$R^2$ model can be found in Refs.~\cite{He2021tachionic, He2019}.

\subsection{The $\delta N$ formalism}

As established in the previous section, the dominant contribution to the non-Gaussian signal in the intermediate-coupling regime arises from the super-horizon sourcing of the curvature perturbation. This physical setup is perfectly suited for the $\delta N$ formalism established by the authors \cite{Sasaki1996} at linear order, and \cite{Lyth2005} used this formalism to express the primordial non-gaussian signal in multifield inlationary models. During inflation, different regions of the universe follow slightly different trajectories in field space, leading to variations in the local expansion each region undergoes. The $\delta N$ formalism captures precisely these differences: regions that expand by a larger amount accumulate a positive curvature perturbation. According to this formalism, evaluating the uniform density perturbation $\zeta$ on super-horizon scales $k \ll aH$ is equivalent to the perturbation of the number of e-foldings $\mathcal{N}(t,t_i,\mathbf{x})$ from an initial flat hypersurface at $t_i$ to a final uniform density hypersurface with $\delta \rho =0$ at $t$. Typically, the initial hypersurface is chosen at th horizon crossing $k = aH$. Hence, at super-horizon scales, the uniform density perturbation can be written as \cite{Sasaki1996}

\beq
    \zeta(t\,,\mathbf{x}) = \mathcal{N}(t, t_i,\mathbf{x}) - N(t, t_i) = \delta N\,,
    \label{delta N }
\enq
where the local expansion $\mathcal{N}(t, t_i,\mathbf{x})$ at some time $t$ is defined as

\beq
    \mathcal{N}(t, t_i,\mathbf{x}) = \int_{t_i}^t H(t',\mathbf{x})\,\dif t'\,,
\enq 
where $H(t',\mathbf{x})$ is the local Hubble expansion rate due to the perturbartions, which depends on the scalar fields $\phi^I$, $H(t, \mathbf{x}) = H(\phi^I(t\,,\mathbf{x}))$ (under the slow-roll approximation, the field velocities $\dot{\phi}^I$ can be neglected). The background, unperturbed expansion $N$ is defined in a similar way

\beq
   N(t, t_i) = \int_{t_i}^t H(t')\,\dif t'\,.
\enq

If we take the initial time $t_i = t_*$ as the time when horizon crossing occurs, $k = aH$, then $\mathcal{N}$ will be a function of $\phi^I$ evaluated at horizon-crossing.
Expanding the number of e-folds $N(\phi^I)$ in a Taylor series around the background field values $\bar{\phi}^I_*$ at horizon crossing time $t_*$, the expression \eqref{delta N } can be expressed as

\begin{equation}
    \zeta(t\,,\mathbf{x}) \simeq  N_I(t)\,\delta\phi^I_*(\mathbf{x}) + \frac{1}{2} N_{IJ}(t)\, \delta\phi^I_*(\mathbf{x}) \delta\phi^J_*(\mathbf{x}) + \dots\,,
    \label{deltaN_expansion}
\end{equation}
where $N_I \equiv \nabla_I N = \partial_I N$ and $N_{IJ} \equiv \nabla_I \nabla_J N $ are the first and second covariant derivatives of the e-fold number with respect to the background field values, respectively, and $\delta\phi^I$ is the covariant gauge-invariant field perturbation given by \eqref{inflationary perturbation} evaluated at $t_*$, and are assumed to be nearly Gaussian. In the Fourier space,  we have 

\beq 
\zeta_\mathbf{k}(t) = N_I(t)\,\delta \phi^I_*(\mathbf{k}) +\dfrac{1}{2}N_{IJ}(t)\,[\delta\phi^I \star \delta \phi^J]_*(\mathbf{k})+\cdots\,,
\label{delta N formalism}
\enq
where $[\delta\phi^I \star \delta \phi^J]_*(\mathbf{k})$ is the convolution product between the field perturbations $\delta \phi^I$. With the relation \eqref{delta N formalism}, the 2-point and 3-point functions of $\zeta$ can be expressed easly as \cite{Seery2005,Vernizzi2006,Gao2008, Elliston2012}

\begin{align}
      \nonumber  \langle \zeta(\mathbf{k}_1)\,\zeta(\mathbf{k}_2) \rangle &= N_I N_J\, \langle \delta\phi^I_*(\mathbf{k}_1)\,\delta \phi^J_*(\mathbf{k}_2) \rangle \\
    &= (2\pi)^2 \,\delta^{(3)}(\mathbf{k}_1 + \mathbf{k}_2)\,P_\zeta(k)\,,
    \label{2-point function}
\end{align}

\begin{multline}
    \langle \zeta(\mathbf{k}_1)\,\zeta(\mathbf{k}_2)\,\zeta(\mathbf{k}_3)\rangle = N_I\, N_J\, N_K\,\langle \delta\phi^I(\mathbf{k}_1)\,\delta\phi^J(\mathbf{k}_2)\,\delta^K(\mathbf{k}_3)\rangle + \\
    + \dfrac{1}{2}N_I\,N_J\,N_{KL}\,\langle \delta\phi^I(\mathbf{k}_1)\,\delta\phi^J(\mathbf{k}_2)\,[\delta\phi^K \star \delta \phi^L]_*(\mathbf{k}_3)\,\rangle + \text{perms} \cdots\,,
    \label{3-point expansion}
\end{multline}
where the 2-point function of field perturbations is given by

\beq 
    \langle \delta\phi^I_*(\mathbf{k}_1)\,\delta \phi^J_*(\mathbf{k}_2) \rangle = (2\pi)^3 \delta^{(3)}(\mathbf{k}_1 + \mathbf{k}_2)\,P^{IJ}_{\phi_*}(k)\,.
    \label{2-point perturbations}
\enq

The 3-point function of field perturbations (the term on the first line in \eqref{3-point expansion}) accounts for primordial interactions near horizon crossing and is evaluated via the In-In formalism \cite{Maldacena2003, Weinberg2005}. For the multifield case, this computation was established by \cite{Seery2005} and later generalized to the covariant representation $Q^I$ by \cite{Elliston2012}

\begin{equation}
     \langle \delta\phi^I(\mathbf{k}_1)\,\delta\phi^J(\mathbf{k}_2)\,\delta^K(\mathbf{k}_3)\rangle =(2\pi)^3\,\delta^{(3)}\left(\sum_i\mathbf{k}_i\right)\dfrac{4\pi^4}{\prod_i k_i^3}|P_{\phi_*}|^2\,\mathcal{A}^{IJK }\,,
   \label{3-point function1}
\end{equation}
where $P_{\phi_*}^2 = H_*^2 /4\pi^2$ is the spectrum of massless scalar fields $\phi^I$ and  $\mathcal{A}^{IJK}$ is a function of $\mathbf{k}_i$, which contains information about the shape and the amplitude of the non-Gaussian signal. In addition, it also contains information about the non-trivial field space geometry, throughout the Riemann tensor $R_{IJKL}$ and its covariant derivatives.
The computation of $\mathcal{A}^{IJK}$ for multifield inflation models was carried out by \cite{Seery2005}, \cite{Lyth2005}, \cite{Gao2008} and \cite{Elliston2012} for a non-flat field space. By employing these results, we couple the initial field non-Gaussianity with the subsequent super-horizon evolution governed by the $\delta N$ coefficients, ensuring that the field space geometry of the Higgs-$R^2$ model is fully accounted for at the moment of exit. 

The next-leading order in \eqref{3-point expansion} can be computed using the Wick theorem, which can be reduced to products of 2-point functions. The 3-point function at next-to-leading order is given by

\begin{multline}
        \langle \delta\phi^I(\mathbf{k}_1)\,\delta\phi^J(\mathbf{k}_2)\,[\delta\phi^K \star \delta \phi^L]_*(\mathbf{k}_3)\,\rangle = (2\pi)^3\,\delta^{(3)}\left(\sum_i \mathbf{k}_i\right) \\
        \times (P^{IK}_{\phi_*}(k_1)\,P^{JL}_{\phi_*}(k_2) + P^{IJ}_{\phi_*}(k_2)\,P^{kL}_{\phi_*}(k_3) + P^{IK}_{\phi_*}(k_3)\,P^{JL}_{\phi_*}(k_1))\,,
        \label{3-point function2}
\end{multline}
where $P^{ab}_{\phi_*}(k_i)$ is the power spectrum  of field perturbations given by \eqref{2-point perturbations}, which is related to the 2-point function of $\zeta$ according to \eqref{2-point function}. The 3-point functions \eqref{3-point function1} and \eqref{3-point function2} directly contribute to the final shape and amplitude of non-Gaussian signal thorugh the primordial bispectrum function $\mathcal{B}_\zeta(k_1\,,k_2\,,k_3)$.


\section{Primordial bispectrum} 
\label{sec4}

In order to connect the 3-point statistic \eqref{3-point expansion} with observations, we quantify the amount of non-Gaussianity using the non-linear parameter $f_\text{NL}$, which encodes the deviation from Gaussianity. In the local form of non-Gaussianity, the curvature perturbation $\zeta(x)$ can be expressed as a local, non-linear expansion around a purely Gaussian background field $\zeta_g(x)$. In position space, this is conventionally written as \cite{Verde2000, Komatsu2001,Boubekeur2006}:

\beq
    \zeta(x) = \zeta_g(x) - \dfrac{3}{5}f_\text{NL}^\text{loc}\,(\zeta_g(x)^2 - \langle\zeta_g^2(x)\rangle)\,,
\enq
where $\zeta_g(x)$ is the Gaussian contribution of $\zeta$. The 3/5 factor arises from the historical convention relating the curvature perturbation to the Bardeen potential $\Phi$ during the matter-dominated era ($\Phi = - \frac{3}{5}\,\zeta$) \cite{Bardeen1980, Bardeen1983}. Physically, this local ansatz reflects a scenario where the non-Gaussian signal is generated on super-horizon scales. In this regime, spatial gradients are negligible, and a long-wavelength background mode effectively modulates the amplitude of short-wavelength fluctuations. When transformed into momentum space to evaluate the bispectrum, this quadratic real-space coupling becomes a convolution. Consequently, the signal reaches its maximum amplitude in the squeezed limit configuration $k_3 \ll k_1\simeq k_2$. As we will show, this squeezed-limit enhancement is consistent with the super-horizon transfer of perturbations from the isocurvature sector to the adiabatic mode induced by transient turns in the field space trajectory of the Higgs-$R^2$ model. 

The non-Gaussian signal is quantified through the bispectrum
$\mathcal{B}_\zeta(k_1,k_2,k_3)$, defined as

\begin{equation}
    \langle  \zeta(\mathbf{k}_1)\,\zeta(\mathbf{k}_2)\,\zeta(\mathbf{k}_3)\rangle = (2\pi)^3\,\delta^{(3)}\left(\sum_i \mathbf{k}_i\right)\,\mathcal{B}_\zeta(k_1\,,k_2\,,k_3)\,.
    \label{3-point function}
\end{equation} 

In general, the bispectrum function can be defined as a sum over all shapes of the non-Gaussianities \cite{Babich2004}

\beq 
    \mathcal{B}_\zeta(k_1\,,k_2\,,k_3) \propto  \sum_\text{type}f_\text{NL}^\text{type}\,S_\text{type}(k_1\,,k_2\,,k_3)\,,
\enq
where $S_{\rm type}(k_1,k_2,k_3)$ denotes the shape function, which characterizes the momentum dependence of the bispectrum and encodes information about the underlying inflationary dynamics. Using the conventional normalization of the bispectrum, the non-linear parameter $f_\text{NL}$ is related to the primordial bispectrum as \cite{Komatsu2001,Maldacena2003} 

\beq 
    f_\text{NL} = \dfrac{5}{6}\,\dfrac{\mathcal{B}_\zeta(k_1\,,k_2\,,k_3)}{P_\zeta(k_1)P_\zeta(k_2) + P_\zeta(k_1)P_\zeta(k_3) + P_\zeta(k_2)P_\zeta(k_3)}\,.
    \label{full f_NL}
\enq 

By taking into account the relations $P_\zeta = N_I\,N_J P^{IJ}_\phi$ and $P^{IJ}_\phi = G^{IJ}\,P_\phi^2$, we can substitute \eqref{3-point function1} and \eqref{3-point function2} into \eqref{3-point expansion} and compare with \eqref{3-point function} to obtain the full expansion for the non-linear parameter $f_\text{NL}$ \cite{Seery2005}:

\begin{equation}
      f_\text{NL} = \dfrac{5}{6}\,\dfrac{N_I\,N_J\,N_K}{(N_I\,N_J\,G^{IJ})^2\,\sum_i k_i^3}\,\mathcal{A}^{IJK}
     + \dfrac{5}{6}\,\dfrac{G^{IK}G^{JM}\,N_I\,N_J\,N_{KM}}{(N_I\,N_J\,G^{IJ})^2} + \cdots\,,
     \label{non-linear parameter}
\end{equation}
where $\mathcal{A}^{IJK} = \mathcal{A}^{IJK}(k_1\,,k_2\,, k_3)$ is the momentum-dependent function introduced in \eqref{3-point function1}. In this expression, the latter term is momentum independent, which corresponds to the usual local form $f_\text{NL}^\text{loc}$ generated by super-horizon evolution \cite{Lyth2005, Vernizzi2006}, whereas the former  encodes intrinsic horizon-crossing interactions that source other non-Gaussian shapes, such as the equilateral ($k_1 = k_2 = k_3 = k_*$) and orthogonal configurations. In this work we evaluate the local, equilateral and folded configurations, with particular emphasis on the local contribution generated by super-horizon mode evolution.

When evaluating Eq.~\eqref{non-linear parameter}, the tensor $\mathcal{A}^{IJK}$ contains the intrinsic field space bispectrum evaluated at horizon crossing. The full exact expression for $\mathcal{A}^{IJK}$, as derived for curved field spaces in Ref.~\cite{Elliston2012}, includes additional terms proportional to the field space Riemann tensor and the second covariant derivatives of the potential. However, these omitted terms primarily generate non-Gaussianity that peaks in the squeezed limit. As established above, the local signal in our model is overwhelmingly dominated by the classical super-horizon evolution encoded in the $N_{IJ}$ terms, rendering the intrinsic local contributions negligible. Therefore, to capture the dominant intrinsic effects for the equilateral configuration, it is sufficient to retain the completely symmetric parts of the derivative interactions. Following \cite{Kaiser2013}, the relevant momentum-dependent contributions for the equilateral shape are:

\begin{equation}
     \mathcal{A}^{IJK} = \sum_{\text{cyclic}}\left[\sqrt{2\epsilon_H}\,\dfrac{\dot{\phi}^I_*}{\dot{\sigma}_*}\,G^{JK}\,\mathcal{F}_1(k_1\,,k_2\,,k_3) + \dfrac{2\epsilon_H}{3}\,G_{AB}\,G_{CD}\dfrac{\dot{\phi^B_*}\,\dot{\phi^D_*}\,}{\dot{\sigma}^2_*}\,\nabla^K R^{IACJ}\,\mathcal{F}_2(k_1\,,k_2\,,k_3) \right]\,,
    \label{A_IJK_expression}
\end{equation}
where $\dot{\phi}^I_*\,,\dot{\sigma}_*$ are evaluated at the moment of horizon crossing and $R^{IACJ}$ is the Riemann tensor of the field space.
The functions $\mathcal{F}_i(k_1,k_2,k_3)$ encode the momentum dependence
of the bispectrum and are given by

\beq
    \mathcal{F}_1(k_1\,,k_2\,,k_3) = \dfrac{k_1(\mathbf{k}_2\cdot\mathbf{k_3})}{2} - 2\dfrac{k_2^2\,k_3^2}{k_1+k_2+k_3}
\enq

\begin{multline}
        \mathcal{F}_2(k_1\,,k_2\,,k_3) = k_1^3\left(N-\log\left(\dfrac{k_1+k_2+k_3}{k_*}\right) -\gamma_\text{euler} - \dfrac{1}{3}  \right) \\
        + \dfrac{4}{9}(k_1 + k_2 + k_3)^3 -(k_1 + k_2 +k_3)\sum_{1<j}k_i\,k_j\,,
\end{multline}
where $\gamma_\text{euler} \approx0.577$ is the Euler-Mascheroni constant. The new contributions \eqref{A_IJK_expression} include a covariant derivative of the Riemann tensor. For a two-dimensional maximally symmetric field space manifold, the Riemann tensor can be written as

\beq
    R_{IJKL} = \dfrac{R_\text{fs}}{2}(G_{IK}\,G_{JL} - G_{IL}\,G_{JK})\,,
\enq
where $R_\text{fs}$ is the field space's  Ricci scalar given in \eqref{Christoffel}. Since the field space manifold of the Higgs-$R^2$ model is maximally symmetric and possesses a constant Ricci scalar \eqref{Christoffel}, the Riemann tensor is covariantly constant, $\nabla^K R^{IACJ}=0$. As a result, the curvature-dependent contribution in Eq.~\eqref{A_IJK_expression} vanishes identically. Thus, only the first term in Eq.~\eqref{A_IJK_expression} contributes to $f_{\rm NL}$.
The cyclic sum corresponds to the permuted terms in pairs of indices and $k_i$. Therefore, $\mathcal{A}^{IJK}$ in the equilateral limit $k_1 = k_2 = k_3 = k_*$ is given by 

\beq
    \mathcal{A}^{IJK} = \dfrac{\sqrt{2\epsilon_H}}{\dot{\sigma}_*}\,\left(\dot{\phi}^I_*\,G^{JK} + \dot{\phi}^J_*\,G^{KI} + \dot{\phi}^K_*\,G^{IJ} \right)\mathcal{F}_1(k_*)\,,
\enq
where $\mathcal{F}_1(k_*) = \left.\mathcal{F}_1(k_1\,,k_2\,,k_3)\right|_\text{equi}$ is the shape function in the equilateral configuration. The momentum conservation condition, $\sum_i \mathbf{k}_i = 0$, implies that the vectors form a closed equilateral triangle. By squaring $\mathbf{k}_1 = -(\mathbf{k}_2 + \mathbf{k}_3)$, the dot product can be expressed strictly in terms of the magnitudes as $\mathbf{k}_2 \cdot \mathbf{k}_3 = -k^2/2$. Substituting this into the shape function yields

\begin{equation}
    \mathcal{F}_1(k_*) = \dfrac{k_*\left(-k_*^2/2\right)}{2} - 2\dfrac{k_*^4}{3k_*} = -\dfrac{11}{12}k_*^3\,.
\end{equation}

These results allow us to express the non-linear parameter $f_\text{NL}$, given by \eqref{non-linear parameter}, as

\beq 
f_\text{NL} = f_\text{NL}^{(3)} + f_\text{NL}^{(4)}\,,
\enq
where

\beq
    f_\text{NL}^{(3)} = \dfrac{5}{6}\,\dfrac{\mathcal{F}_1(k_*)}{k_*^3}\dfrac{1}{(N_I N_J G^{IJ})}\,,
    \label{intri f_NL}
\enq

\beq
    f_\text{NL}^{(4)} = \dfrac{5}{6}\,\dfrac{G^{IK}G^{JM}\,N_I\,N_J\,N_{KM}}{(N_I\,N_J\,G^{IJ})^2}\,.
    \label{local f_NL}
\enq

Our main objective is to evaluate the non-linear parameter $f_\text{NL}$ for the Higgs-$R^2$ model.
While in single-field or strict slow-roll scenarios the derivatives of the number of $e$-folds can
be approximated locally via the gradient of the Hubble parameter
($N_I \approx -H_I / (H \epsilon_H)$), this local approximation breaks down in the presence of
rapid turns in field space. During the evolution of the Higgs-$R^2$ model, the trajectory undergoes
a transient phase in which the dimensionless bending parameter $|\eta_\perp|$ becomes of order
$\mathcal{O}(1)$. Consequently, the field velocity is no longer strictly aligned with the local
gradient, and isocurvature perturbations strongly source the adiabatic mode on super-horizon scales. To capture this dynamics, it is convenient to project the derivatives of $N$ onto the kinematic
basis adapted to the background trajectory, using the unit vectors defined in \eqref{unit vectors}.
The first derivative $N_I$ evaluated at horizon crossing ($t_*$) can be decomposed as

\begin{equation}
    N_I = N_\sigma e_{\sigma I}^* + N_s e_{s I}^*\,,
\label{first derivative N}
\end{equation}
where $N_\sigma$ represents the purely adiabatic contribution, while the
component $N_s$ quantifies the transfer of isocurvature perturbations to the curvature perturbation
$\zeta$ between horizon exit and the end of inflation, and is highly sensitive to the integrated
effect of the turning rate $\eta_\perp$.
 
It is convenient to introduce a unit vector $e_N^I$ parallel to the gradient of $N$, i.e., $e_N^I \propto N^I$, defined in terms of the correlation angle $\Delta$ as \cite{Peterson2011a, Peterson2011}

\begin{equation}
    e_{N I} = \cos \Delta\,e_{\sigma I} + \sin\Delta\,e_{s I}\,.
\label{N flow vector}
\end{equation}

The non-geodesic motion dynamically couples the adiabatic and isocurvature perturbations on super-horizon scales. Consequently, they develop a non-zero cross-correlation by the end of inflation. To quantify this effect, it is convenient to introduce a correlation angle $\Delta$ \cite{Wands2002, Peterson2011}. Following the standard kinematic projection, the cross-correlation coefficient between the primordial curvature and isocurvature perturbations is given by
\begin{equation}
    \sin\Delta = \frac{\mathcal{C}_{\zeta \mathcal{S}}}{\sqrt{\mathcal{P}_{\zeta}\,\mathcal{P}_{\mathcal{S}}}}\,,
\label{cross angle}
\end{equation}
where $\mathcal{P}_\zeta$ and $\mathcal{P}_\mathcal{S}$ are the dimensionless power spectra of the adiabatic and isocurvature modes, respectively, and $\mathcal{C}_{\zeta \mathcal{S}}$ is their cross-correlation spectrum, defined as
\begin{equation}
     \mathcal{P}_\zeta(k) = \frac{k^3}{2\pi^2}\,|\zeta_k|^2, \qquad
     \mathcal{P}_{\mathcal S}(k) = \frac{k^3}{2\pi^2}\,|\mathcal{S}_k|^2, \qquad
     \mathcal{C}_{\zeta\mathcal{S}}(k) = \frac{k^3}{2\pi^2}\,{\rm Re}\!\left(\zeta_k\,\mathcal{S}_k^\ast\right)\,.
     \label{powerspectrum}
\end{equation}
Here, $\mathcal{S}_k = \frac{H}{\dot{\sigma}}\, Q_s$ is the dimensionless isocurvature perturbation.  The presence of a non-zero correlation angle ($\Delta \neq 0$) is a purely multifield feature: the sourcing of $\zeta_k$ by the entropy perturbation $Q_s$ is controlled by the turning rate $\eta_\perp$, as follows from the evolution equation for $\dot{\zeta}_k$. Multifield effects in this model depend directly on the parameter space, particularly on the value of $\xi_h$. Previous analyses of the Higgs-$R^2$ model have shown that sizeable multifield effects arise for $\xi_h \sim \mathcal{O}(10^{-1})$ \cite{Gundhi2020, Pineda2025}. Consequently, the efficiency of super-horizon transfer is sensitive to the specific value of $\xi_h$, and the magnitude of the correlation angle $\Delta$ provides a direct measure of this efficiency.

The definitions \eqref{first derivative N} and \eqref{N flow vector} allow us to rewrite $N_I$ as

\begin{equation}
    N_I = \dfrac{e_{NI}}{\sqrt{2\epsilon_H}\,\cos\Delta}\,,
    \label{N_I vector}
\end{equation}
from which, by comparison with \eqref{first derivative N}, we obtain

\begin{equation}
    N_\sigma = \dfrac{1}{\sqrt{2\epsilon_H}}\,,\qquad
    N_s = \dfrac{\tan \Delta}{\sqrt{2\epsilon_H}}\,.
    \label{N derivaties}
\end{equation}

The second derivative of $N$ is then given by the covariant derivative of $N_I$:

\begin{equation}
    N_{IJ} = \nabla_I N_J = \partial_I N_J - \Gamma^K{}_{IJ} N_K\,.
\label{second derivative N}
\end{equation}

To evaluate the non-linear parameter, we project $N_{IJ}$ onto the kinematic basis, defining the purely adiabatic, mixed, and purely isocurvature second derivatives as $N_{\sigma\sigma} = e_{\sigma *}^I e_{\sigma *}^J N_{IJ}$, $N_{\sigma s} = e_{\sigma *}^I e_{s *}^J N_{IJ}$, and $N_{ss} \equiv e_{s *}^I e_{s *}^J N_{IJ}$, respectively. Additionally, we can simplify the intrinsic contribution by recalling that the definition of the number of $e$-folds, $N_I = \partial N/\partial \phi^I$, yields the background relation $\dot{\phi}^I_* N_I = H_*$. We can obtain closed analytical expressions for these projections using Eqs. \eqref{unit vectors}, \eqref{first derivative N}  and \eqref{N derivaties}. By expanding the covariant derivatives and using the non-zero Christoffel symbols \eqref{Christoffel} of the field space metric, $N_{IJ}$ evaluates to the following expressions:

\begin{multline}
      N_{\sigma \sigma} = \dfrac{1}{\dot{\sigma}^2}\left[\dot{\phi}^2\,\partial_\phi N_\phi + \dot{h}\,\dot{\phi} (\partial_\phi N_h + \partial_h N_\phi) + \dot{h}^2\,\partial_h N_h\right]  + \alpha\,\dfrac{\dot{\phi}\,\dot{h}}{\dot{\sigma}^2}\,N_h - \dfrac{\alpha}{2}\,\dfrac{\dot{h}^2}{\dot{\sigma}^2}\,e^{-\alpha\phi}\,N_\phi  
\end{multline}

\begin{multline}
      N_{ss} = \dfrac{1}{\dot{\sigma}^2}\left[e^{-\alpha\phi}\,\dot{h}^2\,\partial_\phi N_\phi - \dot{h}\,\dot{\phi} (\partial_\phi N_h + \partial_h N_\phi) + e^{\alpha\phi}\,\dot{\phi}^2\,\partial_h N_h\right]  + \alpha\,\dfrac{\dot{\phi}\,\dot{h}}{\dot{\sigma}^2}\,N_h - \dfrac{\alpha}{2}\,\dfrac{\dot{h}^2}{\dot{\sigma}^2}\,N_\phi  
\end{multline}

\begin{multline}
      N_{\sigma s} = \dfrac{1}{\dot{\sigma}^2}\left[e^{-\alpha\phi/2}\,(\dot{\phi}\,\dot{h}\,\partial_\phi N_\phi + \dot{h}^2\partial_\phi N_h) + e^{\alpha\phi/2}\,(\dot{\phi}^2\,\partial_h N_\phi + \dot{\phi}\,\dot{h}\,\partial_h N_h)\right]\\
       -\alpha\dfrac{\dot{\phi}^2}{\dot{\sigma}^2}e^{\alpha\phi/2} N_h - \dfrac{\alpha}{2}\dfrac{\dot{h}\,\dot{\phi}}{\dot{\sigma}^2}e^{-\alpha\phi/2}\,N_\phi\,,
\end{multline}
where the first derivatives $N_\phi$, $N_h$ can be obtained using \eqref{first derivative N} and \eqref{N derivaties}

\begin{align}
    N_\phi &= \dfrac{1}{\sqrt{2\dot{\sigma}^2\,\epsilon_H}}\left(\dot{\phi} + e^{-\alpha\phi/2}\, \dot{h}\tan \Delta\right) \\[0.2cm]
    N_h & = \dfrac{1}{\sqrt{2\dot{\sigma}^2\,\epsilon_H}}\left(\dot{h} + e^{\alpha\phi/2}\,\dot{\phi}\, \tan\Delta\right)\,.
\end{align}

These expressions must be evaluated at horizon crossing $t_*$. We can see that the first derivatives of $N$ can be expressed in terms of the correlation angle $\Delta$, which implies that the cross-correlation spectrum contributes to the enhancement of non-Gaussianities. By substituting this kinematic decomposition into Eqs.~\eqref{intri f_NL} and \eqref{local f_NL}, the non-linearity parameters take the final form:

\begin{equation}
    \dfrac{6}{5}\,f_\text{NL}^{(3)} = -\dfrac{11}{6}\,(\epsilon_H\,\cos^2\Delta)\,,
\end{equation}

\begin{equation}
   \dfrac{6}{5} f_\text{NL}^{(4)} = (2\epsilon_H\,\cos^4\Delta)\,\left(N_{\sigma\sigma} + 2 \tan\Delta \,N_{\sigma s} + \tan^2\Delta\,N_{ss}\right)\,.
\end{equation}

Both contributions to the non-Gaussian signal are suppressed by the slow-roll parameter $\epsilon_H$;  however, the local form contains second derivatives of $N$ in the kinematical basis, which can be large when the energy transfer from isocurvature to adiabatic perturbations on super-horizon scales is efficient. Therefore, the dominant mechanism for non-Gaussianity production in the Higgs-$R^2$ model is isocurvature conversion on super-horizon scales. Consequently, we expect the local contribution $f_{\rm NL}^{\rm loc}$ to dominate the primordial bispectrum whenever the transfer from isocurvature to curvature perturbations is efficient.

Although we have been able to deduce a closed form of $f_\text{NL}^{(4)}$, the second derivatives $N_{\sigma\sigma}$, $N_{ss}$ and $N_{\sigma s}$ requires solving the full multifield dynamics through the transient turning phase. Furthermore, observational configurations probe the total bispectrum rather than its individual physical contributions. Consequently, to capture the exact integrated effect of the isocurvature transfer without relying on local approximations, the full bispectrum $\mathcal{B}(k_1\,,k_2\,,k_3)$ is evaluated numerically using the \texttt{PyTransport} code \cite{pytransport2018}.

\subsection{Numerical results}

With the analytical framework established, we now proceed to the exact numerical evaluation of the non-Gaussian signal. To integrate the full multifield equations of motion and capture the super-horizon isocurvature transfer without relying on slow-roll or local approximations, we implement the model in the \texttt{PyTransport} code \cite{pytransport2018}. We focus our analysis on a fiducial region of the parameter space known to yield viable inflationary trajectories compatible with CMB observations. Specifically, we set the background parameters to:
\begin{equation}
    \xi_h \sim \mathcal{O}(10^{-1})\,,\quad \lambda = 10^{-10}\,,\quad \xi_s = 4\times 10^8\,.
\end{equation}

\begin{figure}[t!]
    \centering
    \includegraphics[width = 0.6\textwidth]{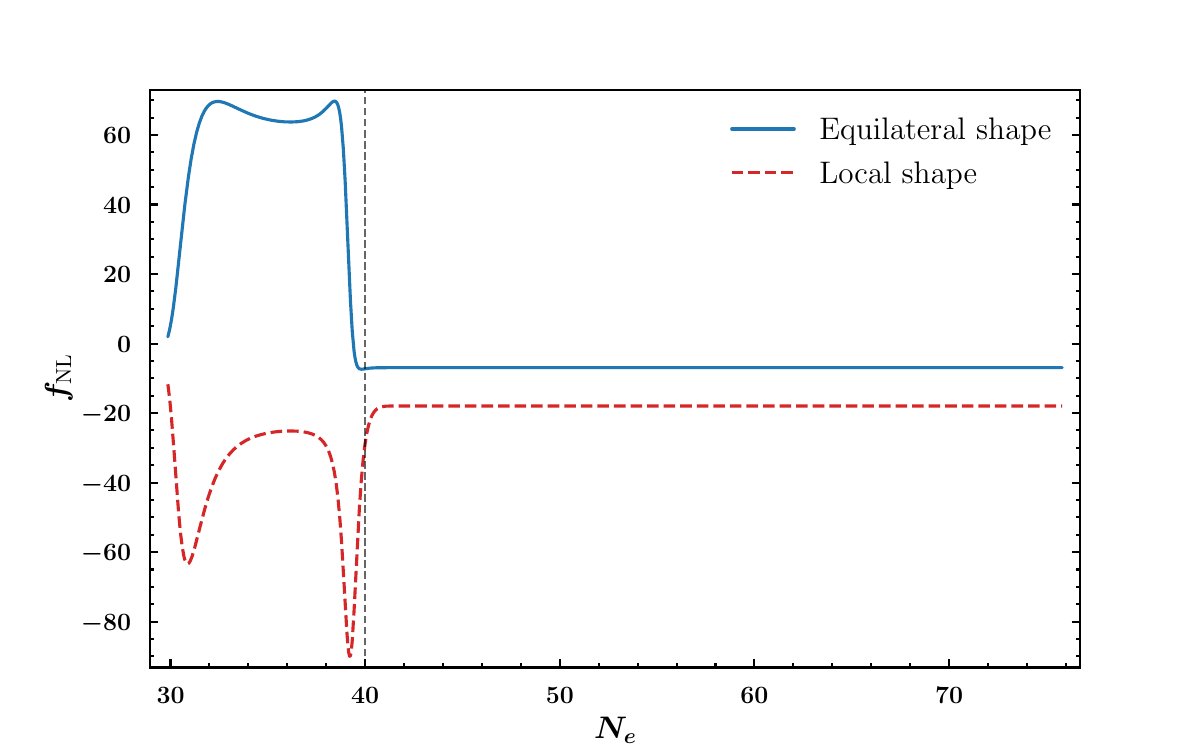}\quad
    \includegraphics[width = 0.55\textwidth]{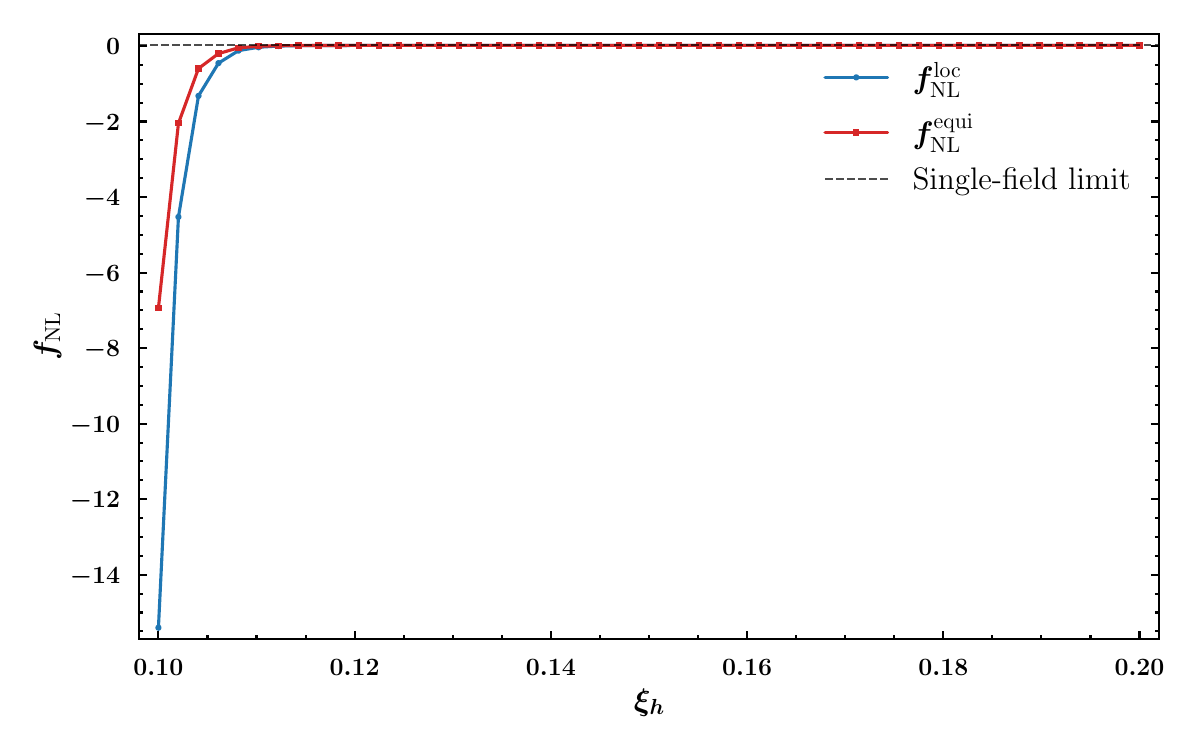}
    \caption{Upper panel: Evolution of the non-linear parameter $f_\text{NL}$ for a nonminimal coupling $\xi_h = 0.1$ in the equilateral and local shapes. Bottom panel: The nonminimal coupling parameter dependence of $f_\text{NL}$. The horizontal dashed line corresponds to the Maldacena  consistency relation value of single field inflation, $f_\text{NL}\to 0.0159$.   }
    \label{fig2}
\end{figure}

In this setup, the Higgs nonminimal coupling $\xi_h$ is treated as the primary free parameter. As we show below (see Fig.~\ref{fig2}), varying $\xi_h$ within this regime directly modifies the geometry of the field space trajectory, thereby governing the efficiency of the isocurvature-to-adiabatic conversion and the final amplitude of $f_\text{NL}$. In particular, for a benchmark value of $\xi_h = 0.1$, the background trajectory undergoes a transient turning phase where the bending parameter becomes $|\eta_\perp| = 0.63$ (see Fig.~\ref{fig1}). This non-trivial geometry strongly couples the curvature and isocurvature perturbations on super-horizon scales, sourcing a non-zero correlation angle $\Delta$ and acting as the main driver for non-Gaussianities. In order to evaluate the full bispectrum, we scan the momentum space by parameterizing the wave numbers $k_1, k_2, k_3$ in terms of the shape parameters $\alpha$ and $\beta$, following \cite{Ronayne2018}:
\begin{align}
    k_1 &= \dfrac{k_*}{2}(1 - \beta)\,, \\
    k_2 &= \dfrac{k_*}{4}(1 + \alpha + \beta)\,, \\
    k_3 &= \dfrac{k_*}{4}(1 - \alpha + \beta)\,,
\end{align}
where $k_*$ is the pivot scale that exits the horizon $N_* = 50$ $e$-folds before the end of inflation. In this parameterization, the physically allowed domain for the spatial triangles is mapped onto a region in the $(\alpha, \beta)$ plane bounded by the vertices $(-1,0)$, $(1,0)$, and $(0,1)$. 

To comprehensively analyze the primordial non-Gaussian signature, we numerically evaluate $f_\text{NL}$ at the three characteristic vertices of this domain: the equilateral, squeezed (local), and folded configurations. 
For the equilateral configuration, the wave numbers are identical ($k_1 = k_2 = k_3 = k_*/3$), corresponding to $(\alpha, \beta) = (0, 1/3)$. 
For the folded configuration, which characterizes non-standard initial states, the triangle flattens into a line ($k_1 = k_2 + k_3$), mapped at $(\alpha, \beta) = (0, 0)$. 
Finally, the local configuration corresponds to the squeezed limit ($k_3 \ll k_1 \simeq k_2$), which is formally reached when $\alpha \to 1$ and $\beta \to 0$. From a numerical standpoint, setting $\alpha = 1$ strictly is prohibitive, as $k_3 = 0$ causes the primordial power spectrum to diverge as $P_\zeta(k_3) \propto k_3^{-3}$. To safely resolve the squeezed limit while maintaining numerical stability, we establish a robust scale hierarchy of $k_3 / k_1 \approx 0.01$ by setting $\alpha = 0.98$ and $\beta = 10^{-3}$. In Fig.~\ref{fig2} we show the evolution of the non-linear parameter $f_\text{NL}$ for equilateral and local shapes (upper panel), and the behaviour of $f_\text{NL}$ as a function of the nonminimal coupling parameter $\xi_h$. (Bottom panel) for the initial conditions $\phi_0 = 5.7$ and $h_0 = 10^{-5}$ in Planck units. Several noteworthy features emerge from the numerical analysis. In the evolution of $f_{\text{NL}}$, we can observe a transient peak during the interval of e-folds where the sharp turning rate occurs (see Fig.~\ref{fig1}). After this turning phase, marked by the dashed line representing its peak, the value stabilizes on super-horizon scales until reaching the final amplitude of $f_{\text{NL}}$ at the end of inflation. In the Bottom panel, we see a strong dependence of the non-linear parameter as $\xi_h$ increases. The asymptotic value corresponds to the single-field limit $f_\text{NL} \to 0.0159$ established by \cite{Maldacena2003}. This behaviour is consistent with the effective single-field limit of the model  \cite{Gundhi2020, He2018, Ema2017} and to the appearence of primordial features in the primordial power spectra \cite{Pineda2025}; for large couplings, the multifield turn is highly suppressed, and the isocurvature perturbations decay efficiently. The model recovers the standard, featureless adiabatic spectrum, and multifield effects disappear.

\begin{figure}[t!]
    \centering
    \includegraphics[width = 0.8\textwidth]{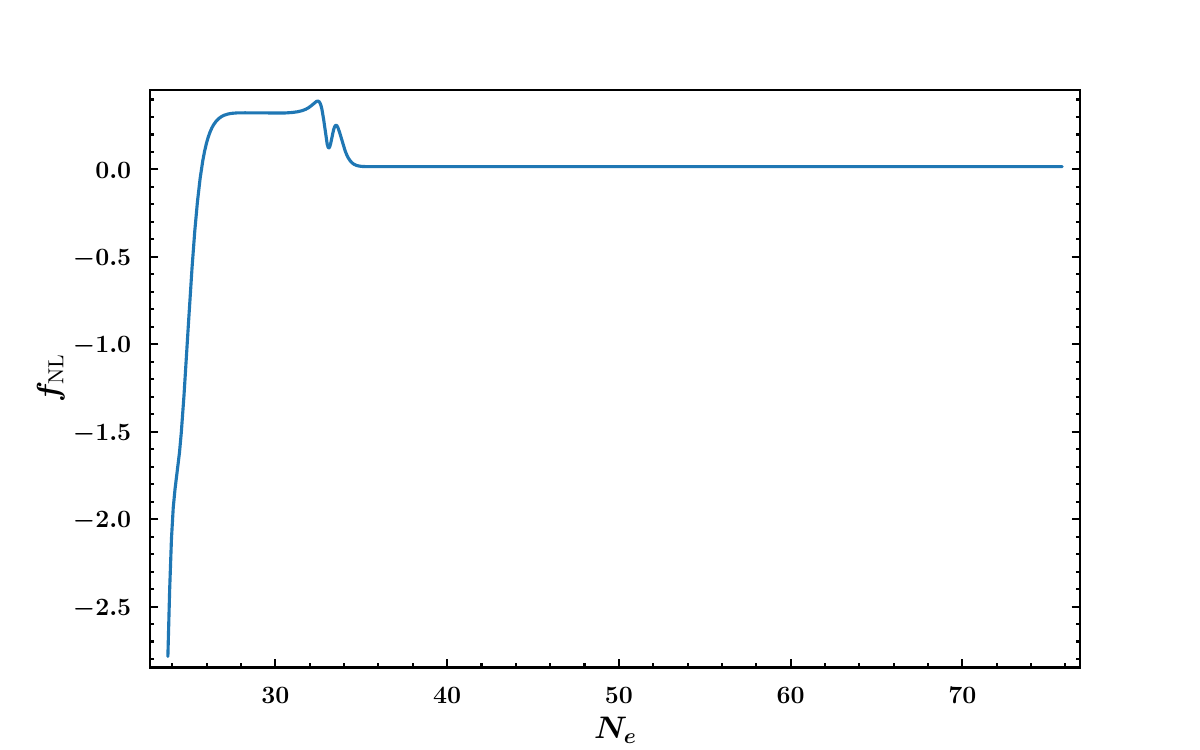}
    \caption{Evolution of $f_\text{NL}$ evaluated on the equilateral triangle  for the pivot scale $k_*$ and for nonminimal coupling parameter $\xi_h = 0.1$ with  Higgs field's initial condition $h_0 = 10^{-4}$ in Planck units. }
    \label{fig3}
\end{figure}

Crucially, the predicted non-Gaussian signal exhibits a strong dependence on both the initial conditions \cite{Kaiser2013,Vernizzi2006,Yokoyama2007,Battefeld2007} and the value of the Higgs nonminimal coupling $\xi_h$. For the fiducial initial condition $h_0 = 10^{-5}$, we find that small values of the coupling lead to sizeable local non-Gaussianities, reaching $f_{\rm NL}^{\rm loc} \simeq -17.7$ at $\xi_h = 0.1$. However, as the nonminimal coupling increases, the amplitude of the bispectrum decreases rapidly and approaches the single-field consistency relation prediction, $f_{\rm NL} \simeq 0.0159$. This behaviour signals a transition from a genuinely multifield regime, where efficient turning trajectories source curvature perturbations, to an effectively single-field attractor in which non-Gaussianities become slow-roll suppressed.

These predictions can be directly confronted with current CMB observations \cite{Planck2018_non_gaussianity}. The large negative values of $f_{\rm NL}^{\rm loc}$ generated at low $\xi_h$ are incompatible with the observational bounds on primordial non-Gaussianity, whereas the effectively single-field regime remains fully consistent with the observed near-Gaussian primordial fluctuations. 
For the fiducial class of trajectories considered in this work, current constraints imply

\begin{equation}
    \xi_h \gtrsim 0.12\,.
\end{equation}

 Our results therefore show that measurements of primordial non-Gaussianity provide a powerful probe of the Higgs nonminimal coupling and can significantly restrict the viable parameter space of the model. The numerical evaluation reveals a clear hierarchy among the standard bispectrum templates. For our benchmark choice $\xi_h = 0.1$, the final amplitudes at the end of inflation $N_\text{end}$ are

\begin{equation}
    f_\text{NL}^\text{loc} = -17.7138\,,  \quad f_\text{NL}^\text{equi} = -8.6729\,.
\end{equation}

\begin{figure}[t!]
    \centering
    \includegraphics[width = .8\textwidth]{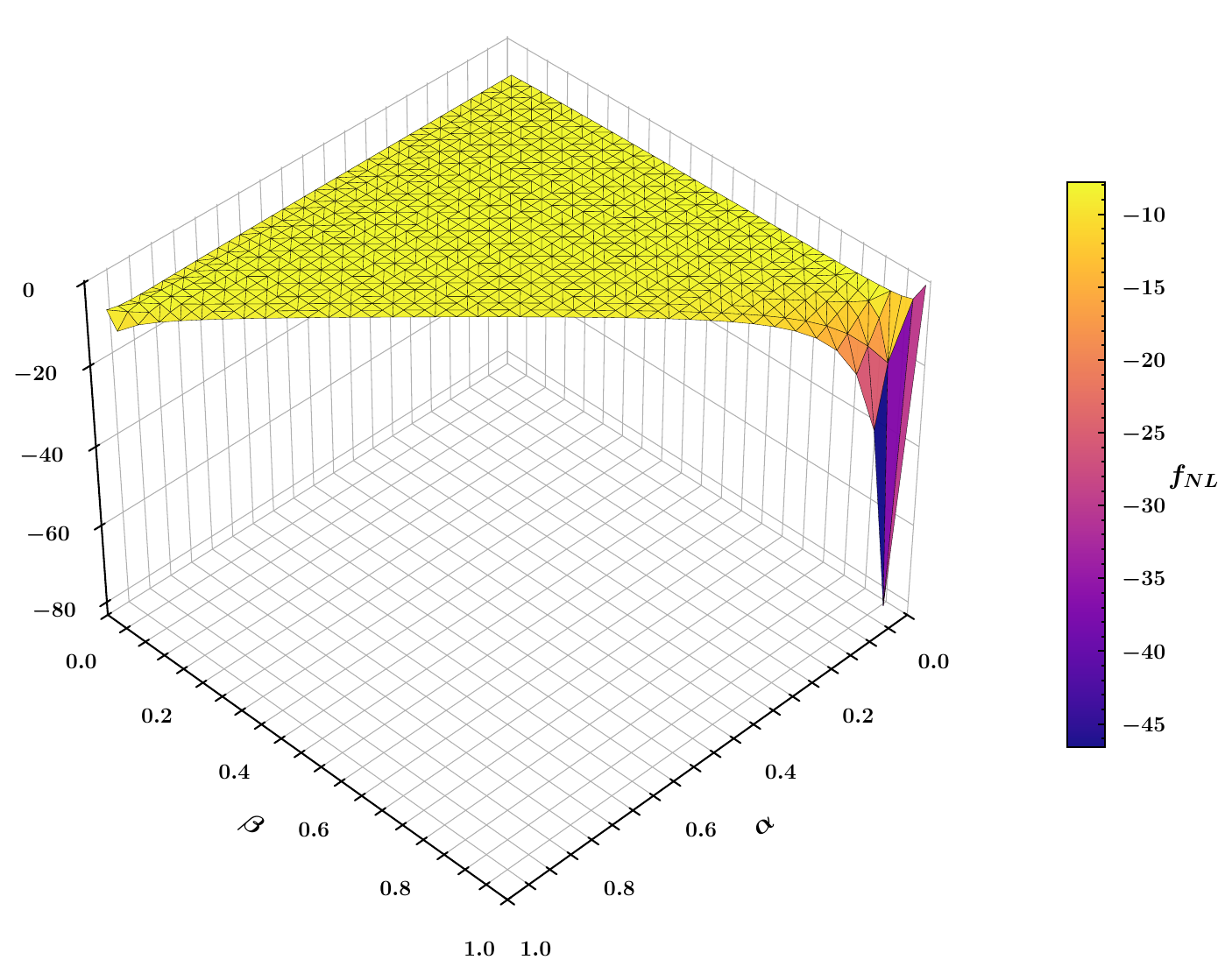}\quad
    \includegraphics[width = .6\textwidth]{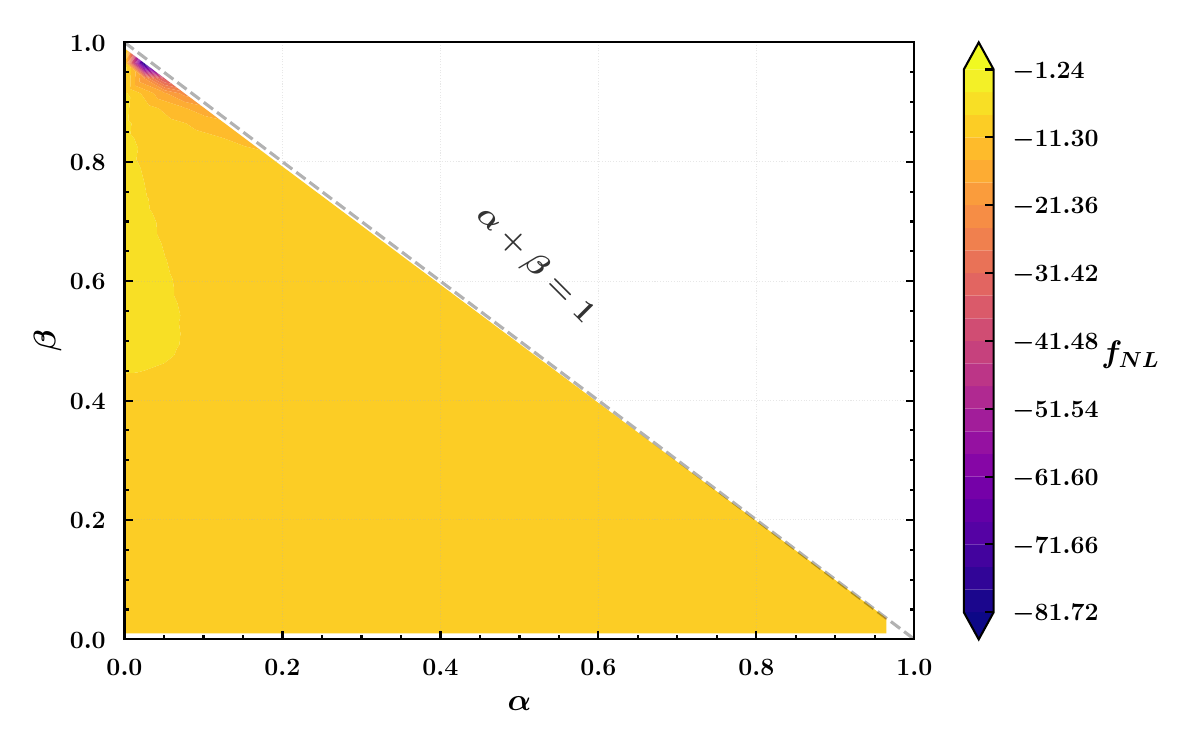}
\caption{\textbf{Shape dependence of the non-linear parameter.} Amplitude of $f_\text{NL}$ in the $(\alpha, \beta)$ plane for the Higgs-$R^2$ model evaluated at the benchmark nonminimal coupling $\xi_h = 0.1$. The physically allowed triangular configurations are bounded by the condition $\alpha + \beta \leq 1$ (dashed line). While the bispectrum amplitude exhibits a relatively flat plateau across the equilateral ($\alpha = 0, \beta = 1/3$) and folded ($\alpha = 0, \beta = 0$) configurations with moderate values, it experiences a sharp negative enhancement towards the vertices of the domain. This drastic divergence at the boundaries corresponds exactly to the squeezed limits (e.g., $k_1 \ll k_2 \simeq k_3$), visually confirming that the primordial non-Gaussianity in this model is overwhelmingly dominated by the local shape driven by superhorizon isocurvature transfer.}
\label{fig4}
\end{figure}

This hierarchical distribution of amplitudes provides sharp insights into the underlying physics. The non-Gaussian signal clearly peaks in the squeezed limit, providing strong evidence that superhorizon isocurvature-to-curvature conversion, described by Eq.~\eqref{local f_NL}, constitutes the dominant mechanism responsible for generating primordial non-Gaussianity in this model. The exact momentum dependence of the non-Gaussian signal is visualized in Fig.~\ref{fig4}, showing the bispectrum amplitude $f_\text{NL}$ across the physically allowed $(\alpha,\beta)$ parameter space. The plots reveal a relatively flat plateau spanning the bulk of the parameter space, encompassing both the equilateral and folded geometries with moderate amplitudes. However, a sharp negative enhancement emerges as the squeezed limit is approached, corresponding to the region $\alpha \rightarrow 0$ and $\beta \rightarrow 1$. This behaviour reflects the rapid growth of the local non-Gaussian signal when one of the momenta becomes much smaller than the other two. This striking feature visually corroborates our analytical findings: the non-Gaussian signal is overwhelmingly dominated by the local shape generated through superhorizon isocurvature-to-curvature conversion.

For completeness regarding the CMB observables, the model consistently yields 

\beq
n_s \simeq 0.969\,,\quad r \simeq 4.93 \times 10^{-3}\,,
\enq
which remain in excellent agreement with the latest Planck  bounds \cite{Pineda2025, Planck2018_inflation, AtacamaDR62025}.

\section{Summary and conclusions}
\label{sec5}

In this work, we have investigated the generation of primordial non-Gaussianities in multifield Higgs--$R^2$ inflation. Using  \texttt{PyTransport} code, we followed the full nonlinear evolution of adiabatic and isocurvature perturbations and computed the primordial bispectrum without relying on slow-roll or local approximations. This approach allowed us to capture the effects of transient turns in the inflationary trajectory and their impact on the statistics of primordial fluctuations. Our analysis shows that the non-Gaussian signal is generated predominantly during a transient turning phase of the background trajectory. When the turning rate becomes significant, isocurvature perturbations efficiently source curvature fluctuations on super-horizon scales, leading to a substantial enhancement of the bispectrum. The resulting signal is strongly dominated by the local configuration, providing clear evidence that superhorizon isocurvature-to-curvature transfer constitutes the primary mechanism responsible for generating primordial non-Gaussianity in this model. A key result of this work is the strong dependence of the non-Gaussian amplitude on the Higgs nonminimal coupling. For moderate values of $\xi_h$, the multifield trajectory undergoes a pronounced turn, generating sizeable local non-Gaussianities that can reach observationally relevant amplitudes. As $\xi_h$ increases, however, the turn becomes progressively suppressed, the isocurvature modes decay more efficiently, and the model approaches an effective single-field attractor. In this regime, the bispectrum amplitude converges to the Maldacena consistency relation, $f_{\rm NL}\rightarrow 0.0159$, and multifield effects become negligible. Comparison with current Planck constraints shows that a significant portion of the low-$\xi_h$ regime is incompatible with observations. For the fiducial initial conditions adopted in this work, the predicted non-Gaussian signal becomes compatible with current observational bounds only for  $\xi_h\gtrsim 0.12$. These results demonstrate that measurements of the bispectrum provide a powerful and complementary probe of Higgs–$R^2$ inflation, capable of constraining aspects of the multifield dynamics that are not accessible through the power spectrum alone.

Future CMB and large-scale structure observations with improved sensitivity to local non-Gaussianity could further test this scenario and provide new information about the geometry of the inflationary field space and the dynamics of multifield inflation.

\section*{Acknowledgments}

This work was partially supported by SECIHTI under grant CBF2023--2024--1937. FP acknowledges financial support from SECIHTI--México under grant No.~803062.

\bibliographystyle{JHEP} 
\bibliography{References}

\end{document}